\documentclass[11pt]{article}       
\usepackage{latexsym}               
\usepackage{url}                    
                                 {\null\hfill$\Box$\par\medskip}
\newtheorem{algX}{Algorithm}
\newenvironment{algorithm}       {\begin{algX}\begin{em}}%
                                 {\par\noindent --- End of Algorithm ---
                                 \end{em}\end{algX}}

\usepackage{graphicx}

\begin{document}


\title{{\bf Exploring the Limits of GPUs\\With Parallel Graph Algorithms}
\footnote{Research partially supported by the Natural Sciences and Engineering Research Council of Canada.}
}

\author{
{\bf Frank~Dehne}\\
{\small School of Computer Science}\\
{\small Carleton University}\\
{\small Ottawa, Canada K1S 5B6}\\
{\small {\em frank@dehne.net}}\\
{\small {\em http://www.dehne.net}}
\and 
{\bf Kumanan Yogaratnam}\\
{\small School of Computer Science}\\
{\small Carleton University}\\
{\small Ottawa, Canada K1S 5B6}\\
{\small {\em kyogarat@connect.carleton.ca}}
} 

\maketitle

\begin{abstract}
In this paper, we explore the limits of graphics processors (GPUs)
for general purpose parallel computing by studying problems that require
highly irregular data access patterns: parallel graph algorithms for
list ranking and connected components. Such graph problems represent
a worst case scenario for coalescing parallel memory accesses on GPUs
which is critical for good GPU performance. Our experimental study
indicates that PRAM algorithms are a good starting point for developing
efficient parallel GPU methods but require non-trivial modifications
to ensure good GPU performance. We present a set of guidelines that
help algorithm designers adapt PRAM graph algorithms for parallel
GPU computation. We point out that the study of parallel graph algorithms
for GPUs is of wider interest for discrete and combinatorial problems
in general because many of these problems require similar irregular
data access patterns.

\end{abstract}

\section{Introduction\label{sec:Introduction}}

Modern graphics processors (\emph{GPU}s) have evolved into highly
parallel and fully programmable architectures. Current many-core GPUs
can contain hundreds of processor cores and can have an astounding
peak performance of up to 1 TFLOP. However, GPUs are known to be hard
to program. Since \emph{coalescing} of parallel memory accesses is
a critical requirement for maximum performance on GPUs, problems that
require irregular data accesses are known to be particularly challenging.
Current general purpose (i.e. non-graphics) GPU applications concentrate
therefore typically on problems that can be solved using fixed and/or
regular data access patterns such as image processing, linear algebra,
physics simulation, signal processing and scientific computing (see
e.g. \cite{gpgpu}). In this paper, we explore the limits of GPU computing
for problems that require irregular data access patterns through an
experimental study of parallel \emph{graph} \emph{algorithms} on GPUs.
Here, we consider list ranking and connected component computation.
Graph problems represent a worst case scenario for coalescing parallel
memory accesses on GPUs and the question of how well parallel graph
algorithms can do on GPUs is of wider interest for discrete and combinatorial
problems in general because many of these problems require similar
irregular data access patterns. We also study how the significant
body of scientific literature on PRAM graph algorithms can be leveraged
to obtain efficient parallel GPU methods.

\subsection*{Parallel \emph{Graph} Algorithms on GPUs}

\begin{figure}[H]
\begin{centering}
\includegraphics[clip,height=2.5in]{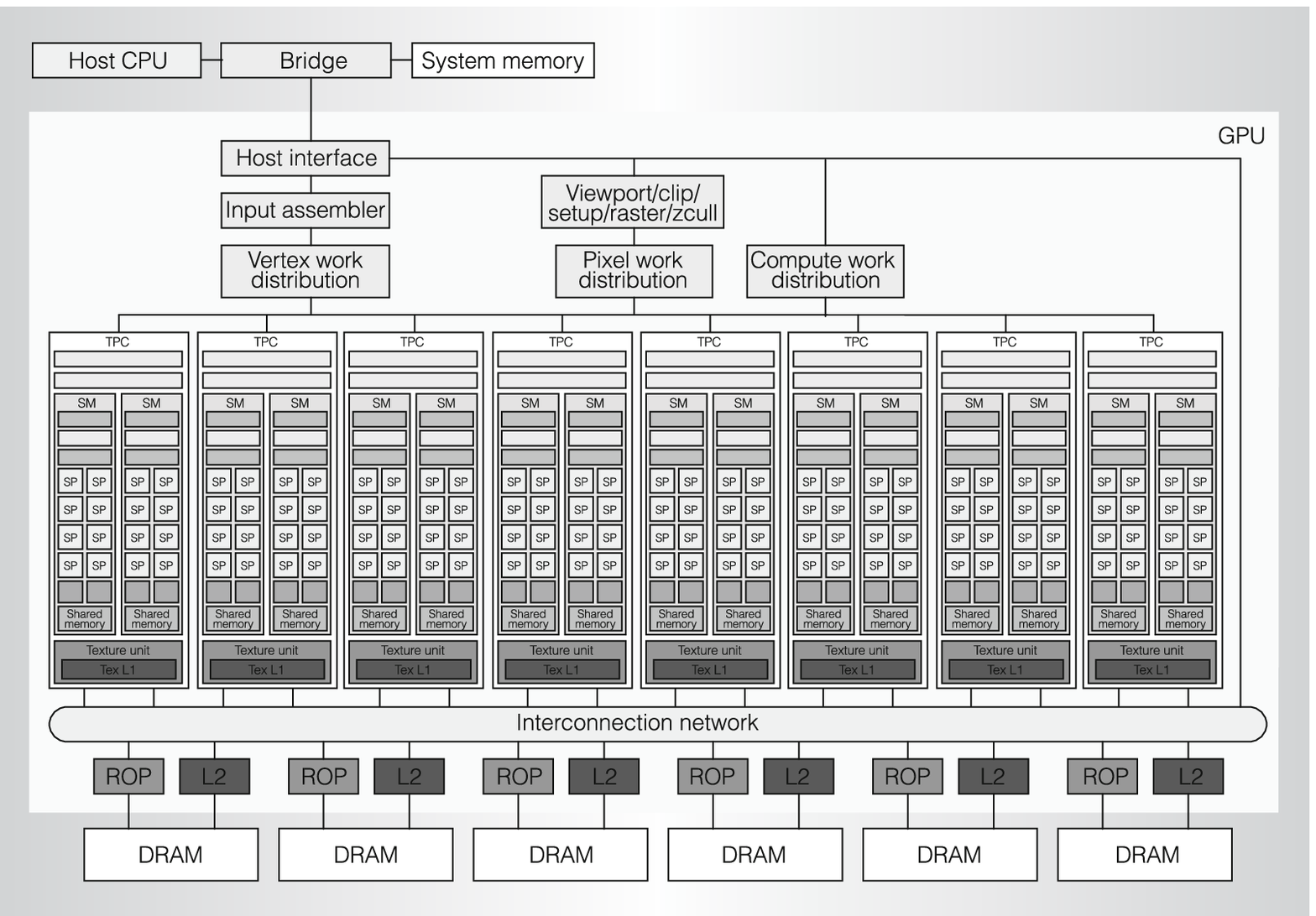} 
\par\end{centering}

\centering{}\caption{\label{fig:nVIDIA-Tesla-Architecture}nVIDIA Tesla Architecture (from\cite{Lindholm2008})}

\end{figure}

In this study, we will focus on nVIDIA's unified graphics and computing
platform for GPUs known as the \emph{Tesla} architecture framework
\cite{Lindholm2008} and associated CUDA programming model \cite{cuda-prog-guide}.
However, our discussion also applies to AMD/ATI's Stream Computing
model \cite{ati-stream-user-guide}and in general to GPUs that follow
the OpenCL standard \cite{opencl-spec,opencl-and-ati-stream-sdk}.
For our experimental study, we used an nVIDIA GeForce 260 with 216
processor cores at 2.1Ghz and 896MB memory. A schematic diagram of
the Tesla unified GPU architecture is shown in Figure \ref{fig:nVIDIA-Tesla-Architecture}.
A Tesla GPU consists of an array of streaming processors (SMs), each
with eight processor cores. The nVIDIA GeForce 260 has 27 SMs for
a total of 216 processor cores. These cores are connected to 896MB
global DRAM memory through an interconnection network. The global
DRAM memory is arranged in independent memory partitions. The interconnection
network routes the read/write memory requests from the processor cores
to the respective memory partitions, and the results back to the cores.
Each memory partition has its own queue for memory requests and arbitrates
among the incoming read/write requests, seeking to maximize DRAM transfer
efficiency by grouping read/write accesses to neighboring memory locations.
Memory latency is optimized when parallel read/write operations can
be grouped into a minimum number of arrays of contiguous memory locations.
GPUs are optimized for streaming data access or fixed pattern data
access such as matrix based operations in scientific computing (e.g.
parallel BLAs \cite{gpgpu}). In addition, their 1 TFLOP peak performance
needs to be matched with a massive need for floating point operations
such as coordinate transformations in graphics applications or floating
point calculations in scientific computing \cite{Lindholm2008}. Parallel
graph algorithms have neither of those two properties. The destinations
of pointers (graph edges) that need to be followed are usually by
definition irregular and not known in advance. The most basic parallel
graph operation, following multiple links in a graph, creates in general
a highly irregular data access pattern. In addition, many graph problems
have no need at all for floating point operations. The question of
how well parallel graph algorithms can do in such a challenging environment
is of wider interest for parallel discrete and combinatorial algorithms
in general because many of them have similar properties.

\subsection*{PRAM vs. GPU}

The \emph{PRAM} model is a widely used \emph{theoretical} model for
parallel algorithm design which has been studied for several decades,
resulting in a rich framework for parallel discrete and combinatorial
algorithms including many parallel graph algorithms (see e.g. \cite{syn-par-alg}).
A PRAM is defined as a collection of synchronous processors executing
in parallel on a single (unbounded) shared memory. PRAMs and GPUs
are similar in that modern GPUs support large numbers of parallel
threads that work concurrently on a single shared memory. In fact,
modern GPUs with 200+ cores \emph{require} large numbers of threads
to optimize latency hiding. An nVIDIA GeForce 260 has a hardware thread
scheduler that is built to manage tens of thousands and even millions
of concurrent threads. The PRAM version most closely related to GPUs
is the CRCW-PRAM supporting concurrent reads and concurrent writes.
Tesla GPUs support concurrent write requests which are aggregated
at each memory partition's memory request queue (using the {}``arbitrary''
model where one of the writes is executed but it is undetermined which
one).\cite{Lindholm2008} In fact, concurrent read/write accesses
are very efficient on GPUs because they nicely coalesce. 

However, GPU and PRAM differ in some important ways. First, as outlined
above, parallel memory requests coming from multiple processor cores
of a GPU need to be coalesced into arrays of contiguous memory locations
\cite{Lindholm2008}. On a PRAM, any set of parallel memory accesses
can be executed in O(1) time, regardless of the access pattern. Second,
as mentioned above, the cores of Tesla GPUs are grouped into streaming
processors (SMs) consisting of eight processor cores each. The cores
within an SM share various components, including the instruction decoder.
Therefore, parallel algorithms for GPUs need to operate SIMD style.
When parallel threads executed on the same SM (and within the same
\emph{warp}, see Section \ref{sec:Optimizing-Parallel-Graph}) encounter
a conditional branch such as an IF-THEN-ELSE statement, some threads
may want to execute the code associated with the {}``true'' condition
and some threads may want to execute the code associated with the
"false" condition. Since the shared instruction
decoder can only handle one branch at a time, different threads can
not execute different branches concurrently and they have to be executed
in sequence, leading to performance degradation. This leads to a need
for SIMD style, data parallel programming of GPUs which is more similar
to classical vector processor programming. Unfortunately, data parallel
solutions for highly irregular problems are known to be challenging. 

We are particularly interested in how efficient parallel graph algorithms
and implementations can be obtained by starting from the respective
PRAM algorithms. Which parts of PRAM algorithms can be transferred
to GPUs and which parts need to be modified, and how?

\subsection*{Summary Of Results}

Our experimental study on parallel \emph{list ranking} and \emph{connected
component} algorithms for GPUs indicates that the PRAM algorithms
are a good starting point for developing GPU methods. However, they
require non-trivial modifications to ensure good GPU performance.
It is critical for the efficiency of GPU methods that parallel data
access coalescing is maximized and that the number of conditional
branching points in the algorithm is minimized. While the number of
parallel threads that can be executed concurrently on a GPU is large,
it is still significantly smaller than the number of PRAM processors.
It is important for the efficiency of GPU methods to appropriately
assign groups of PRAM processors to GPU threads and choose an appropriate
data layout such that GPU threads access data in a pattern referred
to as \emph{striding} (see Section \ref{sub:CUDA-Striding} for precise
definition). Another important difference between the PRAM and GPUs
is that PRAM methods assume zero synchronization overhead while this
is not the case for GPUs. Therefore, the number of actually necessary
and implemented synchronization points for the GPU implementation
needs to be minimized to ensure good performance. The observed GPU
performance for parallel list ranking and connected components appears
to be very sensitive to the total work of the underlying PRAM method.
GPU performance also appears to be more sensitive to the constants
in the time complexity of the algorithm than parallel implementations
for standard multi-core CPUs. This is because GPUs support so many
more threads than multi-core CPUs, each with much less work than a
thread for the corresponding multi-core CPU method, that the constants
in the time complexity have a much larger relative impact. 

For the list ranking problem, the parallel random splitter PRAM method
(Algorithm \ref{alg:Reid-Miller's-Adaptation} below) proposed by
Reid-Miller (and then adapted for the Cray C90) \cite{list-rank-cray-c90}
appears to be a good starting point for an efficient parallel GPU
list ranking method, mainly because it ensures linear total work.
We observed that the parts of our code with irregular access patterns
(following the list pointers) were the dominating parts with respect
to the run time of the entire method. Minimizing the number of data
access for these parts, for example through packing of variables and
caching in GPU registers was crucial for performance. Reid-Miller's
parallel random splitter method is a \emph{randomized} PRAM algorithm
and we observed that the the large number of threads that can be executed
concurrently on a GPU is very helpful to efficiently implement such
methods. The GPU's hardware scheduler was very helpful to make the
implementation surprisingly efficient even when the random selection
of splitter elements created considerable fluctuations in sub-list
lengths. We also observed an interesting inflection point in the running
time curve (as function of list size), where the irregular data access
pattern and data access volume starts to push the limits of the GPU's
on-chip interconnection network. Michael Garland at nVIDIA \cite{Garland-Discussion}
has used our code to reproduce this effect on their machines and pointed
out that this is the first time they have seen such an inflection
point. 

For the connected component problem, we implemented a GPU adaptation
of Shiloach and Vishkin's CRCW-PRAM algorithm \cite{conn-comp-PRAM-OlogN}
following the same guidelines outlined above. Despite the fact that
Shiloach and Vishkin's CRCW-PRAM algorithm requires $O((n+m)\log n)$
work for $n$ vertices and $m$ edges, and the sequential method requires
only linear work, our parallel GPU implementation was significantly
faster than the sequential implementation on a standard sequential
CPU. We also analyzed some interesting performance variations of the
GPU algorithm when executed for different types of graphs.

\section{Implementing Parallel Graph Algorithms On GPUs\label{sec:Optimizing-Parallel-Graph}}

In this section, we discuss general issues regarding the design and
implementation of efficient parallel graph algorithms on GPUs. We
are particularly interested in how efficient parallel graph algorithms
can be obtained by starting from the respective PRAM algorithms and
then modifying them for efficient execution on GPUs.

The Tesla unified GPU architecture supports software development through
the CUDA programming model for GPUs \cite{cuda-prog-guide} which
is an extension of C/C++. A CUDA program executes serial code on the
CPU which then calls a sequence of \emph{kernels} that are executed
on the GPU. As discussed in Section \ref{sec:Introduction}, a Tesla
GPU consists of an array of streaming processors (SMs), each with
eight processor cores. CUDA programs are aware of the hardware in
that each kernel consists of a set of parallel threads that are grouped
into \emph{blocks}. The threads within each block are executed on
the same SM and are grouped into \emph{warps} consisting of 32 threads
each. Each \emph{warp} is executed SIMD style. A hardware thread scheduler
schedules all blocks over all available SMs. Synchronization is available
for all threads within a block. Synchronization across different blocks
requires barrier synchronization \cite{Lindholm2008}. It is worth
noting that there is no stack in the Tesla GPU architecture and hence
there is no recursion supported for the threads within CUDA kernels.

\emph{Global} memory, also called \emph{device} memory, is located
off the GPU chip and on the GPU card similar to memory on a motherboard
for regular CPUs. All global memory locations are accessible by all
cores of the GPU through the on-chip interconnection network that
routes and schedules all accesses to global memory. Global memory
is arranged in independent memory partitions. Each memory partition
has its own queue for memory requests and arbitrates among the incoming
read/write requests, seeking to maximize DRAM transfer efficiency
by grouping read/write accesses to neighboring memory locations. For
general purpose GPU computations, the input data is typically transferred
from the CPU's memory to the GPU's global memory, and after the GPU
has finished its computation, the result is transferred back from
the GPU's global memory to the CPU's memory.

In the remainder of this section we discuss issues that need to be
addressed when designing efficient parallel GPU graph algorithms by
starting from the respective PRAM algorithms and then modifying them
for efficient execution on GPUs.

\subsection{Coalescing Global Memory Accesses\label{sub:Coalescing-Global-Memory}}

When using global memory, a critical requirement for obtaining good
performance is to coalesce memory accesses performed concurrently
by different threads. The goal is to combine multiple global memory
access requests executed concurrently by multiple threads into one
single memory transaction for one of the independent memory partitions
of global memory. The performance improvement that can be gained through
coalescing of memory accesses can be substantial \cite{cuda-prog-guide}.
Our experiments indicate that coalesced data access can improve the
total run time of a CUDA kernel by an order of magnitude (see e.g.
Table \ref{tab:Kernel-Timings-LR-ParRandSplit}).

The GPU's hardware and system support for coalescing memory accesses
has been changing over time. On older systems (CUDA 1.1), memory access
coalescing required the correct alignment of memory accesses in the
algorithm \cite[p.81-88]{cuda-prog-guide}. Each thread needed to
access consecutive memory addresses relative to thread order. For
example, four threads $T_{0}$through $T_{3}$ needed to access memory
addresses $A_{0}$ through $A_{3}$ such that $A_{0}<A_{1}<A_{2}<A_{3}$
and $A_{1}-A_{0}=A_{2}-A_{1}=A_{3}-A_{2}$. Memory accesses where
coalesced in half-warps (16 processor cores) whereby sixteen consecutive
32 bit reads become a single 64 byte memory access transaction. This
fixed coalescing scheme was rather inflexible and complicated to handle.
Newer systems (starting with CUDA 1.2) are more flexible in that concurrent
global memory accesses within half-warps (16 processor cores) that
fall inside the same memory segment whose size is governed by the
size of the memory accesses can be coalesced. Table \ref{tab:CUDA-1.2-Global-Mem-Segments}
shows the segment sizes for different memory accesses issued by half-warps
\cite{cuda-prog-guide}. For example, if 16 data accesses of 2 Bytes
each fit exactly into a 32 Bytes memory segment then this creates
one memory transaction of 32 Bytes. If those 16 data accesses are
not adjacent but fit into a 64 Bytes memory segment then this creates
one memory transaction of 64 Bytes. If the transactions fall into
multiple segments then this creates multiple memory transactions which
is bad for performance and should be avoided. Note that, for data
access sizes beyond the 8 Bytes data size shown in Table \ref{tab:CUDA-1.2-Global-Mem-Segments},
multiple 128 Bytes coalesced memory transactions will be issued.

\subsection{Concurrent Write Memory Accesses\label{sub:Concurrent-Write-Memory}}

CUDA supports concurrent write attempts to global memory without causing
any failure in the execution. As discussed, the entire global memory
is accessible by all cores of the GPU through the on-chip interconnection
network that routes all memory requests to the respective memory partitions.
Each memory partition has its own queue for memory requests and arbitrates
among the incoming read/write requests. Concurrent writes are executed
in \emph{arbitrary} order resulting in one of them succeeding while
the others are effectively ignored. Concurrent read requests are all
handled by the same memory partition in one single memory transaction.
Our experiments show that concurrent reads from the same address are
extremely fast; faster than parallel reads on different memory locations.
A mix of concurrent reads and writes for the same memory location
is not recommended since they will be executed in arbitrary order,
resulting in race conditions.

\subsection{SIMD/SIMT Thread Execution\label{sub:CUDA-Threading} }

A GPU application does not generally need to be aware of the number
of cores. It can create thousands or millions of threads, as needed.
All threads are divided into blocks of up to 768 threads, and each
block is executed by an SM consisting of eight processor cores. A
hardware scheduler performs the assignment of blocks to SMs. An SM
executes a thread block by breaking it into groups of 32 threads called
\emph{warps} and executing them in parallel using its eight cores.
More precisely, the SM performs context switching between the different
warps. This allows the SM to hide the latency of memory access operations
performed by the threads and provides increased computational performance.
When a warp is being executed, the eight cores also perform context
switching between the warp's 32 threads. As indicated in Section \ref{sec:Introduction},
the eight cores of an SM share various hardware components, including
the instruction decoder. Therefore, the threads of a warp are executed
in SIMT (single instruction, multiple threads) mode, which is a slightly
more flexible version of the standard SIMD (single instruction, multiple
data) mode. The active threads of a warp all need to execute the same
instruction as in SIMD mode while operating on their own data. The
main problem arises when the threads encounter a conditional branch
such as an IF-THEN-ELSE statement. Depending on their data, some threads
may want to execute the code associated with the "true"
condition and some threads may want to execute the code associated
with the "false" condition. Since the
shared instruction decoder can only handle one branch at a time, different
threads can not execute different branches concurrently and they have
to be executed in sequence, leading to performance degradation. The
SMs provide a small improvement through an instruction cache that
is shared by the eight cores. This allows for a "small"
deviation between the instructions carried out by the different cores.
For example, if an IF-THEN-ELSE statement is short enough so that
both conditional branches fit into the instruction cache then both
branches can be executed fully in parallel. However, a poorly designed
algorithm with too many and/or large conditional branches can result
in serial execution and very low performance.

\subsection{Thread Synchronization\label{sub:CUDA-Synchronization}}

The PRAM model assumes full synchronization at the level of individual
steps and does not account for synchronization overhead; i.e. synchronization
is without cost. On real multiprocessors including GPUs this is of
course not the case. CUDA supports two types of synchronization. Threads
within a thread block can be synchronized by calling \texttt{\_\_syncthreads()}
from inside the kernel. Threads in different thread blocks can not
be synchronized within the same kernel. Barrier synchronization across
all threads and thread blocks is achieved by breaking an algorithm
into a sequence of different kernels along the barrier synchronization
boundaries. When mapping PRAM algorithms to GPUs, the algorithm designer
needs to be aware that thread synchronization is costly. The PRAM
algorithm needs to be examined and only the absolutely necessary synchronizations
should be implemented. For good performance, the number of synchronization
points needs to be minimized.

\subsection{Striding \& Partitioning\label{sub:CUDA-Striding}}

The PRAM model typically assumes one parallel thread per data item.
This is in many cases not possible for GPUs, even though we are allowed
to allocate millions of concurrent threads. Current nVIDIA Tesla architectures
provide up to 4 Gigabyte global memory which may exceed the number
of threads that a GPU can handle. Also, even though the hardware thread
scheduler is very efficient, there is a cost associated with thread
scheduling and an excessive number of threads leads to performance
degradation. In other cases there are also algorithmic reasons for
having fewer threads than data items; see e.g. the Parallel Random
Splitter List Ranking method in Section \ref{sub:Parallel-Random-Splitter}.
Hence, when implementing e.g. Shiloach and Vishkin's PRAM connected
component algorithm \cite{conn-comp-PRAM-OlogN,syn-par-alg} for graphs
of size $N$ on a GPU using $p$ threads, we need to assume that $N>p$.
Consider $N$ data items in an array $A[0],$..., $A[N-1]$ that need
to be accessed by $p$ threads $T_{0}$, ..., $T_{p-1}$. We distinguish
between two types of access patterns.\emph{ }(1) \emph{Striding:}
Thread $T_{i}$ accesses data item $A[i+s\cdot p]$ in step $s$ for
$0\leq s<\frac{N}{p}$ and $0\leq i\leq p-1$.\emph{ }(2) \emph{Partitioning}:
Thread $T_{i}$ accesses data item $A[i\frac{N}{p}+s]$ in step $s$
for $0\leq s<\frac{N}{p}$ and $0\leq i\leq p-1$.

On standard multi-core CPUs with multiple fully functional processor
cores that have several caching levels for memory access, \emph{partitioning}
usually provides the best performance by making sure that each processor
gets its own chunk of data to access in sequence without needing to
refill its cache. On a many-core GPU where the SMs contain multiple
SIMD cores with a memory access system that favors coalesced memory
accesses across parallel warps and half-warps, \emph{striding} provides
better performance because it optimizes coalescing of memory accesses.
Hence, when porting PRAM algorithms to GPUs, it is important to chose
a data layout that supports striding.

\section{Parallel List Ranking On A GPU}

A basic operation required by nearly all (parallel or sequential)
graph algorithms is to traverse a linked list of edges/pointers. In
this section, we will therefore start with the classical parallel
\emph{list ranking} problem (see e.g. \cite[p. 80]{syn-par-alg})
and study how to convert well known PRAM algorithms for list ranking
into efficient GPU implementations by following the guidelines outlined
in the previous Section \ref{sec:Optimizing-Parallel-Graph}.

Consider a linked list of length $n$ represented by an array $succ\left[0..n-1\right]$
where each $succ[i]$ points to the next element in the linked list.
The first element of the linked list is $succ[0]$ and the last element
has the property $succ[j]=j$. The ranks of all list elements (distances
to the last element of the list) are reported as an array $rank\left[0..n-1\right]$.
We will first study a GPU implementation of the straightforward pointer
jumping algorithm for the PRAM, also known as \emph{Wylie's Algorithm}
\cite[p. 64]{syn-par-alg} \cite{wylie-lr-79}. This algorithm does
provide some limited speedup but, similar to a preliminary result
in \cite{list-rank-CUDA-weak}, suffers from the fact that it requires
$O\left(n\log n\right)$ work. GPU performance appears to be very
sensitive to the total work and it is therefor critical to base the
GPU method on an algorithm with $O\left(n\right)$ work. Cole and
Vishkin's \emph{deterministic coin tossing} method \cite{cole-vish-86}
provides an optimal PRAM list ranking method with $O\left(n\right)$
work but is so complicated and has such high constant factors that
it's performance gain would only materialize for unreasonably high
data sizes far beyond the memory size of a GPU. Many other parallel
list ranking algorithms have been studied in the literature (see e.g.
\cite{sibeyn-1,Sibeyn-2,Dehne-Song,syn-par-alg,list-rank-cray-c90}).
A well suited PRAM algorithm adaptation for our purposes is the \emph{parallel
random splitter} algorithm presented by Reid-Miller for parallel list
ranking on a Cray C90\cite{list-rank-cray-c90}. The Cray C90 is a
vector processor and shares some features with GPUs such as SIMD style
data parallelism. 

In the remainder of this section we present the issues encountered
and results obtained when porting these two methods, Wylie's algorithm
and Reid-Miller's parallel random splitter\emph{ }algorithm, to an
nVIDIA GPU. As a baseline for comparison, we also implemented both
methods on a standard quad-core CPU.

\subsection{Implementing Wylie's Algorithm On A GPU\label{sub:Implementing-Wylie's-Algorithm}}

Wylie's algorithm \cite[p. 64]{syn-par-alg} \cite{wylie-lr-79} is
the simplest parallel algorithm for list ranking. Algorithm \ref{alg:Ptr-Jump-Par-List-Ranking-p<n}
shows a high level GPU pseudo code. In the following we discuss our
GPU adaptations implemented on top of the straight PRAM method implementation.

The GPU pseudo code shown in Algorithm 1 shows two kernels, one for
initialization and one for pointer jumping. It is possible to implement
this in one single kernel. The restriction however is that the kernel
can only be executed using a single thread block. For multiple thread
blocks, barrier synchronization requires the use of multiple kernels
as discussed in Section \ref{sub:CUDA-Synchronization}. The single
kernel implementation is faster for smaller data sets because it can
use the faster \texttt{\_\_syncthreads()} function for synchronization
(see Section \ref{sub:CUDA-Synchronization}). We will make use of
this method in the parallel random splitter list ranking method discussed
in Section \ref{sub:Parallel-Random-Splitter}. For large data sets
(linked lists), we require a multi kernel implementation. Our multi
kernel implementation consists of an outer loop that is executed on
the CPU and calls first an initialization kernel and then a sequence
of $\log\left(n\right)$pointer jumping kernels. Our implementation
went through various optimizations that improve the performance of
this PRAM method on a GPU. The code assumes $n>p$ list elements and
implements \emph{striding} as outlined in Section \ref{sub:CUDA-Striding}
which optimizes coalescing of memory accesses. We made efforts to
avoid conditional branching (e.g. IF-THEN-ELSE statements) through
the use of arithmetic and boolean statements that have the same effect
but avoid the SIMD performance penalty associated with conditional
branching. Another improvement was obtained by clustering data accesses
within the code and assigning the values to registers rather than
making these data accesses throughout the code. This decreases the
number of global memory accesses and improves performance. As discussed
in Section \ref{sub:Coalescing-Global-Memory}, up to 128 bytes can
be read or written as a single transaction (CUDA 1.2). We introduced
a 64 Bit "union" structure which combines
two 32 Bit variables into one. The kernels themselves use the union
data type as a register, and to manipulate the values in register
memory. Each value is read only once from, and written once to, global
memory as a combined single 64bit operation. The use of 64 Bits to
encode two 32 Bit variable allowed us to fully leverage memory coalescing.

\subsection{Parallel Random Splitter List Ranking\label{sub:Parallel-Random-Splitter}}

A PRAM algorithm adaptation presented by Reid-Miller for parallel
list ranking on a Cray C90\cite{list-rank-cray-c90} is the \emph{parallel
random splitter} algorithm; see Algorithm \ref{alg:Reid-Miller's-Adaptation}.
For $r$ sub-lists, Algorithm \ref{alg:Reid-Miller's-Adaptation}
requires $O\left(n+r\lg r\right)$ work and runs in $O\left(\frac{n}{p}+\lg r\right)$
expected time with very small constants. Algorithm \ref{alg:Rand-split-par-list-rank-p=00003Dr}
shows the pseudo code for our GPU adaptation of Reid-Miller's algorithm.
Our implementation uses $p=r$ threads, thus selecting and ranking
$p=r$ splitters. The algorithm requires $O\left(n+p\lg p\right)$
work and $O\left(\frac{n}{p}+\lg p\right)$ expected time. The work
is $O\left(n\right)$ if $p\lg p\leq n$. For $n=1,000,000$ we can
therefore use up to $p\leq62,500$ to ensure $O\left(n\right)$ work.
Parallel list ranking is only useful for large linked lists and $n$
will usually be larger than $1,000,000$ in practice. In our experiments
we found that for optimal performance $p$ is best chosen to be a
small factor times the number of physical processor cores on the GPU.
Even for high values of $n$ we typically used a fixed value for $p$
that is considerably smaller than the maximum allowed to maintain
linear work.

\begin{algorithm}\label{alg:Reid-Miller's-Adaptation}
{\bf Parallel Random Splitter Algorithm \cite{list-rank-cray-c90}}

\begin{enumerate}
\item {\footnotesize Randomly divide the list into $r$ sub-lists by randomly
choosing $r$ }\emph{\footnotesize splitter}{\footnotesize{} nodes.
Reduce each sub-list to a single node with value equal to the number
of values in the sub-list. Now the list is of length $r$. }{\footnotesize \par}
\item {\footnotesize Find the list ranks of the reduced list of splitter
nodes selected in Step 1 using Wylie's algorithm }\cite[p. 64]{syn-par-alg}
\cite{wylie-lr-79}{\footnotesize . These values are the final ranks
of the $r$ splitter nodes.}{\footnotesize \par}
\item {\footnotesize Expand the nodes in the reduced list back into the
original linked list filling in the rank values along the list.}
\end{enumerate}

\end{algorithm}

Our GPU adaptation of Reid-Miller's algorithm shown in Algorithm \ref{alg:Rand-split-par-list-rank-p=00003Dr}
consists of five kernels. Kernel \emph{RS1} in Algorithm \ref{alg:Rand-split-par-list-rank-p=00003Dr}
initializes the supporting data structure which is an \emph{owner}
link for each node, referring to the splitter heading the sub-list
containing the node. Initially, each node starts without a link to
it's owning splitter. Kernel \emph{RS2} in Algorithm \ref{alg:Rand-split-par-list-rank-p=00003Dr}
select the random splitters and sets each splitter's owner link to
point to itself. For random number generation we used the KISS algorithm
by Marsaglia and Zaman\cite{KISS-rand-93} which has a very high period
of $2^{123}$ while using straight-forward 64 bit integer operations
that make it a good match for the GPU architecture and CUDA system.
We also used the KISS algorithm to generate the input data for our
experimental evaluation (Section \ref{sub:Experimental-Results-List-Ranking}).
Kernel \emph{RS3} in Algorithm \ref{alg:Rand-split-par-list-rank-p=00003Dr}
traverses the sub lists for each splitter, counting ranks relative
to each sub list until a node with a different owner (splitter) is
encountered. After the walk, we store each splitter's sub list length
and index as part of a separate short linked list of splitters. Kernel
\emph{RS4} in Algorithm \ref{alg:Rand-split-par-list-rank-p=00003Dr}
ranks the linked list of splitters with the stored sub list lengths
as each splitter's initial rank, using the single kernel implementation
of Wylie's Algorithm outlined in Section \ref{sub:Implementing-Wylie's-Algorithm}.
Kernel \emph{RS5} in Algorithm \ref{alg:Rand-split-par-list-rank-p=00003Dr}
calculates the final rank of each node using its associated splitter's
rank computed in the previous step.

\subsubsection*{48Bit vs. 64Bit Packing Schemes For "Mark"
and "Rank" Arrays}

Our GPU implementation of Algorithm \ref{alg:Rand-split-par-list-rank-p=00003Dr}
uses a 16 bit "mark" array and a 32 bit
"rank" array for marking ownership and
ranking of each linked list node, respectively. The linked list itself
is represented by a 32 bit array of successor links. We also implemented
an alternate version where "mark" and
"rank" are packed into one single 64 bit
value. We will refer to these two implementations as the 48 bit and
64 bit versions of Algorithm \ref{alg:Rand-split-par-list-rank-p=00003Dr}.
One key difference between the two versions is that in the 48 bit
versions the ownership mark is restricted to 16 bit and thus the ranking
algorithm cannot be invoked with more than 16,384 threads. In our
experiments, this did not pose a limitation since 16,384 threads correspond
to 64 thread blocks, which in the case of the GTX 260 GPU is at least
three times the number of physical SMs. For future systems with more
SMs, this could however become a limiting factor.

An important difference between the two versions is in their performance.
Packing the ownership and rank into a single 64 bit value allowed
the implementation to perform a single read and write instead of the
two reads and writes needed for the 48 bit version. As illustrated
in Figure \ref{fig:Comparing-CUDA-LR-Performance}, the 64 bit version
(in green) clearly out-performs the 48 bit version for lists with
up to approximately 52 million nodes, after which the 48-bit version
is the better performer. See Section \ref{sub:Experimental-Results-List-Ranking}
for more details.

Our GPU implementation of Wylie's Algorithm discussed in Section \ref{sub:Implementing-Wylie's-Algorithm}
used a packed 64 bit mark and rank data structure and we observed
that our implementation required extra host memory. The 64 bit version
of our parallel random splitter (Algorithm \ref{alg:Rand-split-par-list-rank-p=00003Dr})
implementation avoids this by re-using the \texttt{succ\_d} array
that was allocated in device memory to supply the linked list data
to return the ranking. This is accomplished by calling the aggregation
kernel with the \texttt{succ\_d} array as the \texttt{rank\_d} parameter,
and then copying the contents of that array into the host's rank array.

\subsubsection*{Slow vs. Fast Kernels}

Table \ref{tab:Kernel-Timings-LR-ParRandSplit} shows running times
of our GPU implementation of Algorithm \ref{alg:Rand-split-par-list-rank-p=00003Dr}
for various values of $n$ and both the 48 bit and 64 bit packing
schemes discussed above. Table \ref{tab:Kernel-Timings-LR-ParRandSplit}
also shows for the running times of the individual kernels in Algorithm
\ref{alg:Rand-split-par-list-rank-p=00003Dr}. It is interesting to
note that while kernels RS3 and RS5 both perform $O\left(n\right)$
work, RS3 requires more than an order of magnitude more running time
than RS5. This is a good example of best and worst case scenarios
for \emph{coalescing} of parallel memory accesses. In RS3, each thread
selects its splitter node and traverses the list from that node onwards
until it encounters a node that belongs to a different splitter/thread.
Since the linked lists used in our experiments are completely random,
every step leads the thread to access some random new position and
there is little opportunity for memory coalescing. On the other hand,
in RS5 each thread does not follow the linked list and is instead
striding over the nodes in the linked list in array order, subtracting
each node's local rank from its associated splitter's rank. The memory
access pattern is ideal for memory access coalescing. As discussed
in \ref{sub:Coalescing-Global-Memory}, this leads to significant
performance improvement.

\subsubsection*{Random Splitter Distribution For The GPU}

As noted in \cite{list-rank-cray-c90}, the parallel random splitter
method implementation for the Cray C90 performed well because the
number of threads used was considerably larger than the number of
actual processors. This is an important requirement for randomized
algorithms to utilize the law of large numbers and perform well. Fortunately,
as discussed in Section \ref{sec:Introduction}, a GPU can handle
thousands and up to millions of threads and this is helpful for implementing
randomized PRAM algorithms on GPUs such as the parallel random splitter
algorithm in \cite{list-rank-cray-c90}. This observation is illustrated
in Table \ref{tab:Comp-Perf-Random-Even-Splitters}. We compare the
sub-list length distribution and kernel performance of our Algorithm
\ref{alg:Rand-split-par-list-rank-p=00003Dr} implementation with
a modified implementation of Algorithm \ref{alg:Rand-split-par-list-rank-p=00003Dr}
where we provide a perfect set of splitters that partition the linked
list into exactly equal size sub-lists. We also show the expected
sub-list length according to the formula shown in \cite{list-rank-cray-c90}.
Table \ref{tab:Comp-Perf-Random-Even-Splitters} shows that the GPU
implementation follows exactly the predicted values in \cite{list-rank-cray-c90}
due to the large number of threads that we are able to instantiate.
Table \ref{tab:Comp-Perf-Random-Even-Splitters} also shows that the
running time of Algorithm \ref{alg:Rand-split-par-list-rank-p=00003Dr}
with perfect even splitters is only marginally better than the running
time of Algorithm \ref{alg:Rand-split-par-list-rank-p=00003Dr} with
random splitters. This is somewhat surprising since the maximum sub-list
length for random splitters shown in Table \ref{tab:Comp-Perf-Random-Even-Splitters}
can be an order of magnitude larger than the maximum sub-list length
for perfect even splitters. Here, our GPU implementation benefits
again from the large number of threads and hardware thread scheduling
on the GPU. Due to the large number of sub-lists, as long as more
sub-lists than processor cores are still active, all of the GPU hardware
is still in full use. The GPU does the load balancing automatically
through its hardware thread scheduling. This is a very interesting
feature of GPUs and we expect this to be in general very helpful for
porting randomized PRAM algorithms for discrete and combinatorial
problems to GPUs.

\subsection{Run Time Comparison For List Ranking\label{sub:Experimental-Results-List-Ranking}}

\begin{figure}[H]
\centering{}\includegraphics[height=6.5cm]{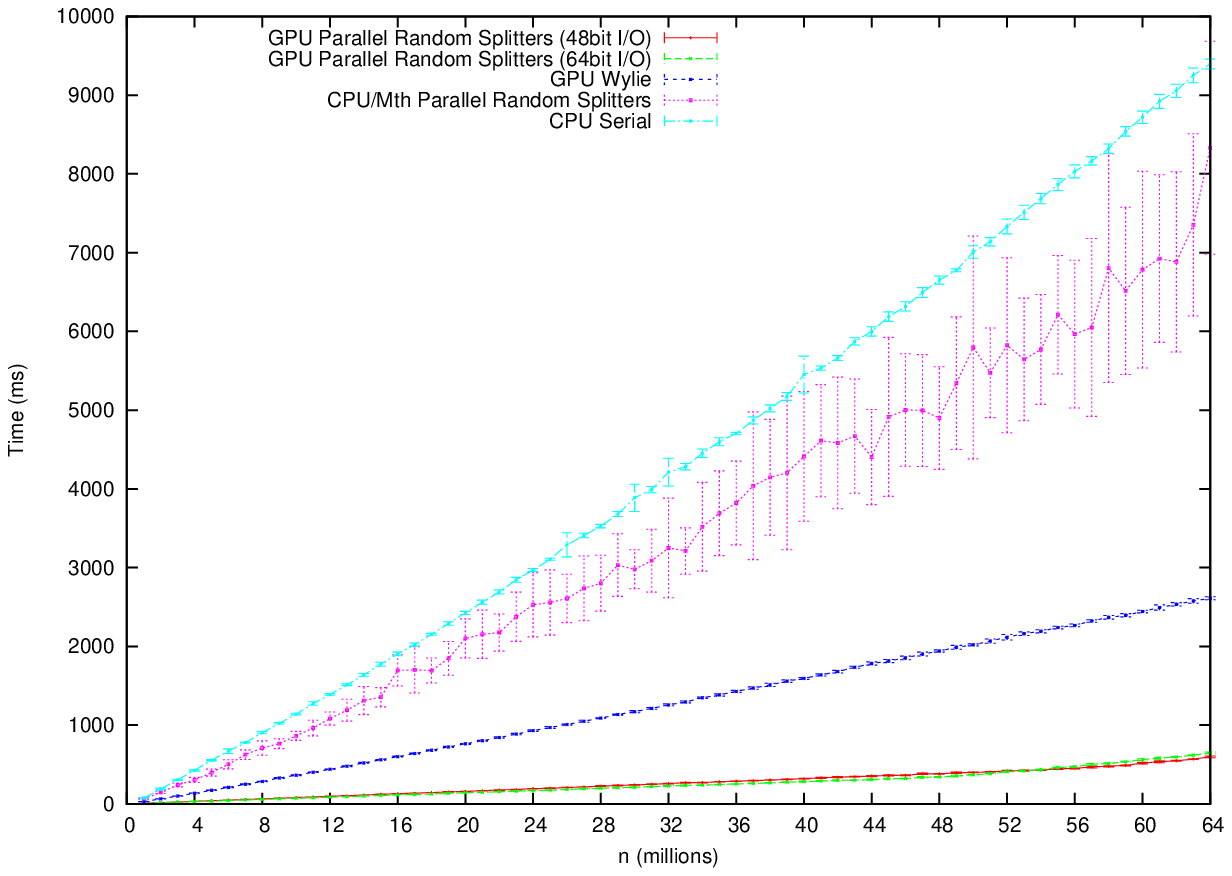}\caption{\label{fig:Comparing-All-List-Ranking}Comparing The Performance of
CPU and GPU List-Ranking Implementations}

\end{figure}

Figure \ref{fig:Comparing-All-List-Ranking} shows a comparison of
the run times (in milliseconds), as a function of list size, for all
of our parallel list ranking implementations: sequential list ranking
on a CPU (Intel Core 2 Quad with 8 GB memory running Fedora Core Linux),
multi threaded list ranking on the same CPU, our GPU implementation
of Wylie's algorithm and our GPU implementation of the parallel random
splitter algorithm (48 bit and 64 versions). The GPU times were measured
on an nVIDIA GeForce 260 with 27 SMs (216 processor cores). Each data
point represents the average of 20 experiments and the vertical bars
represent standard deviation. In general, it appears that our GPU
implementations of PRAM methods seem to be reasonably successful.
Despite the highly irregular data accesses which complicate memory
access coalescing, our GPU list ranking implementation appears to
be a factor 20 faster than sequential list ranking on a standard CPU
which is consistent with the general notion that one SM of an nVIDIA
GeForce 260 GPU is approximately as fast as one standard CPU core
\cite{gpgpu}.

Note that, the multi threaded list ranking on the CPU also uses a
parallel random splitter approach but with much fewer threads since
it has only four cores. The smaller number of threads leads to more
fluctuation in sub list length which is reflected in more fluctuation
in running time as shown in Figure \ref{fig:Comparing-All-List-Ranking}.
Furthermore, the random memory accesses caused by the random linked
list appear to be causing many cache misses on the CPU which create
additional fluctuations in running time.

\begin{figure}[H]
\centering{}\includegraphics[width=0.45\columnwidth,height=5cm]{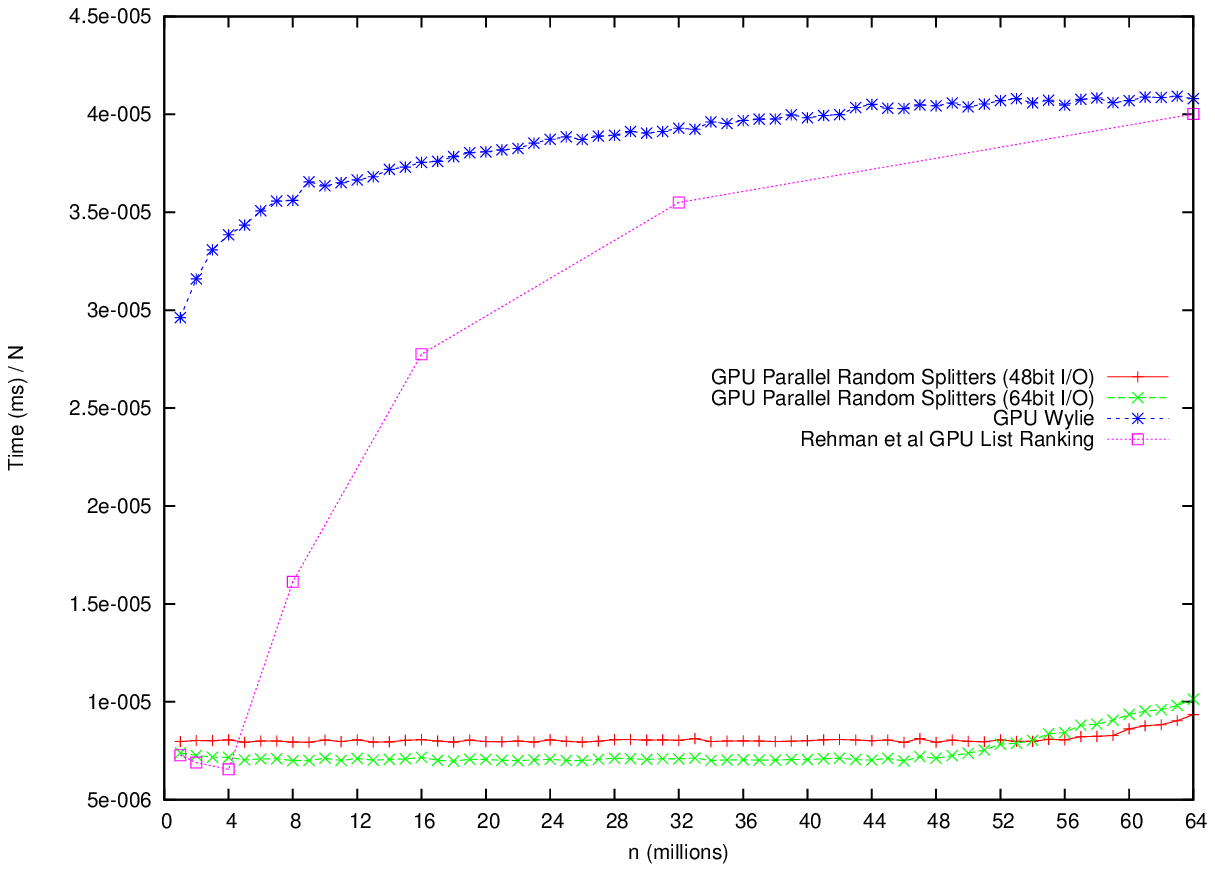}
\includegraphics[width=0.45\columnwidth,height=5cm]{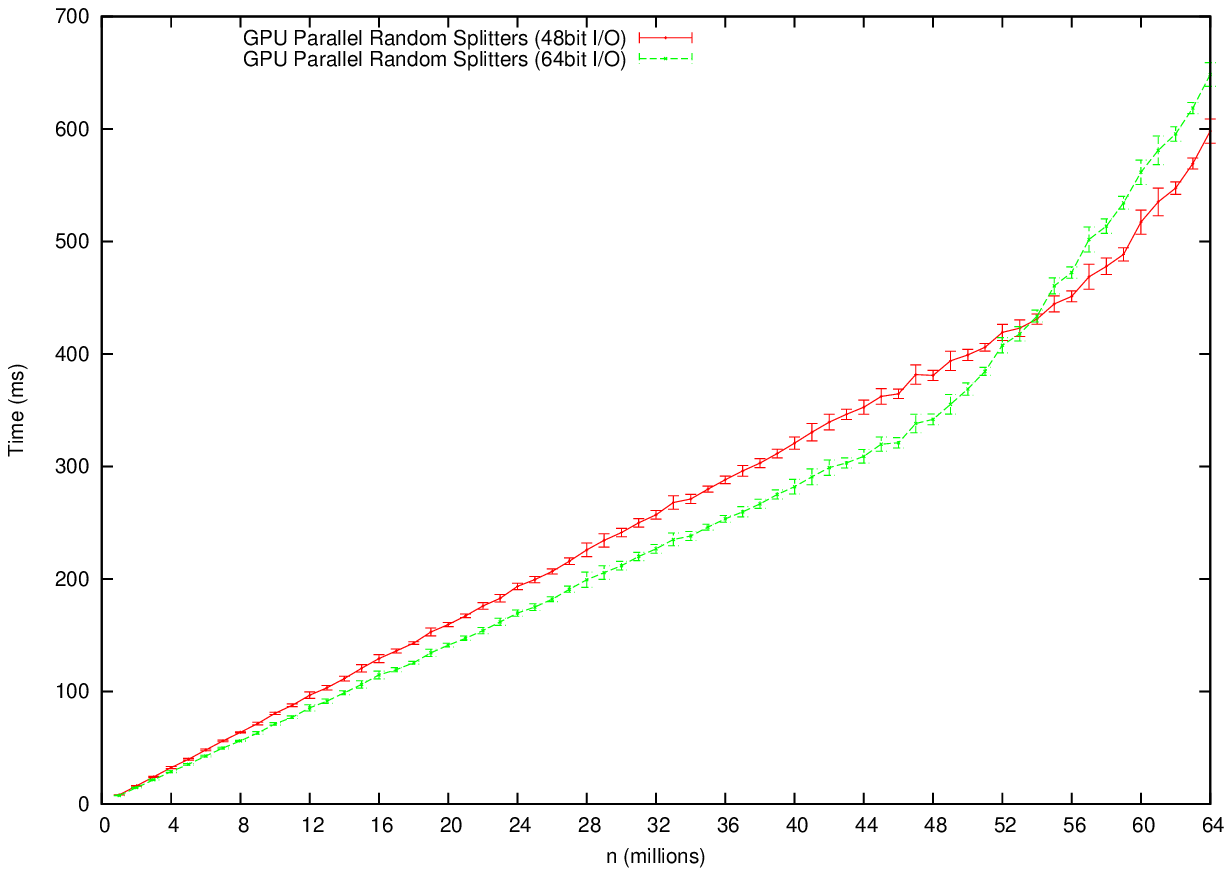}\caption{\label{fig:Comparing-CUDA-LR-Performance}Comparing The Performance
of GPU List Ranking Implementations}

\end{figure}

The left diagram in Figure \ref{fig:Comparing-CUDA-LR-Performance}
shows a more detailed view of the same data for our GPU implementations.
The x-axis represents again list size but the y-axis shows time per
list element (rather than absolute time). Our GPU Wylie and random
splitter implementations both show very little overhead as list sizes
increase. In fact both show a running time growing at nearly the same
rate as data size. In comparison, the time per list element of a modified
Wylie method for GPUs presented earlier by Rehman et al \cite{list-rank-CUDA-weak}
shows a significant increase as data size grows, indicating a significant
growth in overhead per data element with increasing list size. Most
importantly, the diagram highlights how much our random splitter method
implementation for the GPU outperforms Wylie's method and Rehman et
al's method\cite{list-rank-CUDA-weak}. The main reason is that both
Wylie's and Rehman et al's methods require $O\left(n\log n\right)$
work whereas the random splitter method requires $O\left(n\right)$
work. The $\frac{time}{n}$ curves shown are indeed $O\left(\log n\right)$
for Wylie's and Rehman et al's methods, and $O\left(1\right)$ for
our GPU implementation of the random splitter method.

The right diagram in Figure \ref{fig:Comparing-CUDA-LR-Performance}
shows the absolute running times for the 48 bit and 64 bit versions
of our parallel random splitter GPU implementations (Algorithm \ref{alg:Rand-split-par-list-rank-p=00003Dr})
in more detail. We observe an interesting crossing point that occurs
around $n\cong54,000,000$. The 64 bit version of Algorithm \ref{alg:Rand-split-par-list-rank-p=00003Dr}
out-performs the 48 bit variation until $n\cong54,000,000$. For larger
$n$, the 48 bit version is faster. We observe an inflection in the
running timing for the 64 bit version at $n\cong46000000$. The inflection
is a result of the the work starting to overload the memory access
network bandwidth. The bottleneck results in increased time as $n$
increases beyond $54,000,000$. The 48 bit variation of the algorithm
has an inflection point as well, but it occurs later at $n\cong58000000$.
Both inflection points are the result of an overload of the memory
access network on the GPU. Here, the parallel graph algorithm is indeed
testing the limits of the GPU, largely because of the irregular data
access that is typical for PRAM graph methods and many other parallel
discrete and combinatorial algorithms. More precisely, in each iteration
for Kernel RS3 of Algorithm \ref{alg:Rand-split-par-list-rank-p=00003Dr},
the 48 bit version of the kernel requires a 16 bit write, a 32 bit
write, a 32 bit read and a 16 bit read for every list node, and these
accesses are randomly distributed in global memory which is not helpful
for data access coalescing. Each iteration for RS3 issues approx. $n$
memory access transactions moving a total of $96 n$ bits
of memory. In the 64 bit variation, each iteration for RS3 issues
a 64 bit write, a 32 bit read and a 64 bit read, which results in approx.
$n$ memory access transactions moving a total of
$160  n$ bits of memory. Both versions of Algorithm \ref{alg:Rand-split-par-list-rank-p=00003Dr}
start to overload the memory access network for different values of
$n$. 

We discussed our observation with Michael Garland at nVIDIA \cite{Garland-Discussion}.
He requested a copy of our code and he was able to reproduce the same
effect on their machines. Michael Garland pointed out that this inflection
effect had not been observed before since GPU applications typically
have regular data access patterns. The irregular data access patterns
generated by parallel list ranking seem to be exploring the limits
of the nVIDIA GPU. Note that the inflection points are only observable
for algorithms with efficient computation times. For example, the
GPU implementation of Wylie's algorithm and the implementation in
\cite{list-rank-CUDA-weak} do not show inflection points simply because
they are inefficient and require $O\left(n\log n\right)$ work.

\section{Parallel Connected Component Computation}

We now turn our attention to parallel graph connected component computation
on a GPU. For a graph $G$ with $n$ vertices and $m$ edges, the
CRCW-PRAM algorithm by Shiloach and Vishkin\cite{conn-comp-PRAM-OlogN,syn-par-alg}
computes connected components in $O\left(\log n\right)$ time with
$O\left((n+m)\log n\right)$ work using $O(n+m)$ processors. The
algorithm assumes that PRAM concurrent writes are implemented using
the ``arbitrary'' model. Algorithm \ref{alg:Shiloach-Vishkin-ConComp-anyP}
shows an outline of our GPU implementation of Shiloach and Vishkin's
CRCW-PRAM algorithm. In the remainder of this section, we outline
some of the important aspects of our GPU implementation of Shiloach
and Vishkin's CRCW-PRAM algorithm and discuss the performance achieved
by our implementation.

\subsubsection*{Use Of Striding \emph{And} Concurrent Writes}

The PRAM algorithm requires $m+n$ processors, one for each edge and
vertex. As discussed in Section \ref{sub:CUDA-Striding}, a GPU can
execute a large number of concurrent threads but not as many as one
thread per data item. In order to improve performance, our CUDA code
implements the striding access pattern outlined in Section \ref{sub:CUDA-Striding}
using $p\leq m$ threads. In addition, Shiloach and Vishkin's CRCW-PRAM
algorithm requires concurrent write operations which is in principle
no problem for a GPU as discussed in Section \ref{sub:Concurrent-Write-Memory}.
However, when both striding \emph{and} concurrent writes are implemented
together, special care must be taken to ensure that the correct semantics
are maintained and race conditions avoided. For example, in Step 1
of Shiloach and Vishkin's algorithm (\cite[p. 60]{conn-comp-PRAM-OlogN})
two actions take place: ``short-cutting'' followed by ``marking''.
In a CUDA implementation using both striding \emph{and} concurrent
writes, this step will execute incorrectly unless the two actions
are separated into two kernels with a barrier synchronization between
them because the marking step relies on the short-cutting step to
be complete for all vertices. Therefore, our GPU implementation shown
in Algorithm \ref{alg:Shiloach-Vishkin-ConComp-anyP} implements Step
1 of Shiloach and Vishkin's algorithm (\cite[p. 60]{conn-comp-PRAM-OlogN})
as two separate kernels labeled ``Step 1a'' and ``Step 1b''.

Step 5 of Shiloach and Vishkin's algorithm (\cite[p. 60]{conn-comp-PRAM-OlogN})
checks whether the previous execution of Steps 1b and 2 resulted in
any changes. In our GPU implementation, this is also implemented through
a combination of both striding and concurrent writes. Each thread
first checks whether any of its items got changed and then all those
threads that detected a change attempt to update a global variable
$w$, thus implementing a parallel ``OR'' through a concurrent write.

\subsubsection*{Memory Access Optimization}

As outlined in Section \ref{sub:Coalescing-Global-Memory}, the GPU's
hardware optimizes the use of bandwidth to global memory through coalescing
of memory accesses. Our GPU implementation of Shiloach and Vishkin's
CRCW-PRAM algorithm makes efforts to save memory bandwidth through
pre-fetching to local registers. This ensures that within each kernel
the same global memory address is accessed only once. Kernels SV1b,
SV2 and SV3 apply this optimization. Table \ref{tab:Counting-the-RW-In-CCKernels}
shows the number of global memory reads and writes performed by each
kernel.

\subsubsection*{Run Time Comparison For Different Types Of Graphs: Lists, Trees,
and Random Graphs}

\begin{figure}[H]
\begin{centering}
\includegraphics[width=0.45\columnwidth]{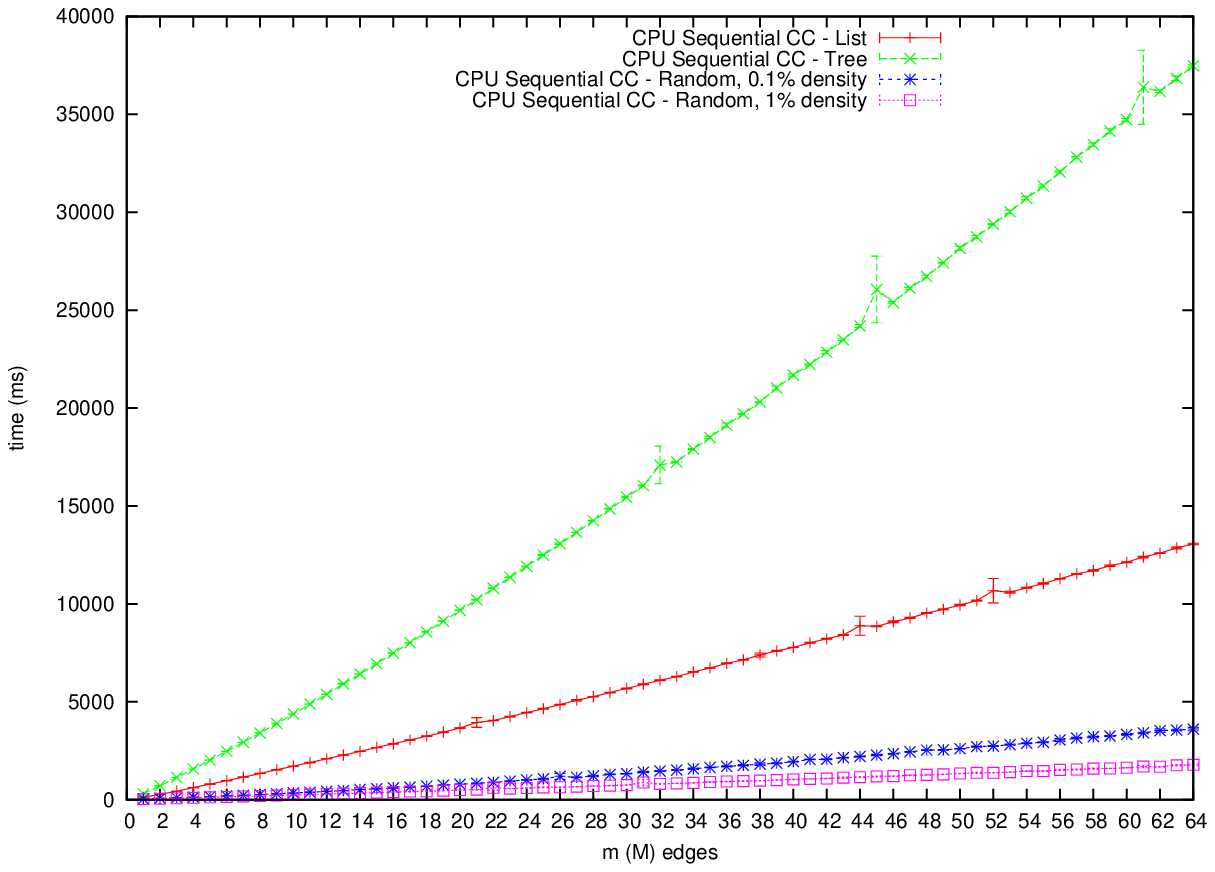}\includegraphics[width=0.45\columnwidth]{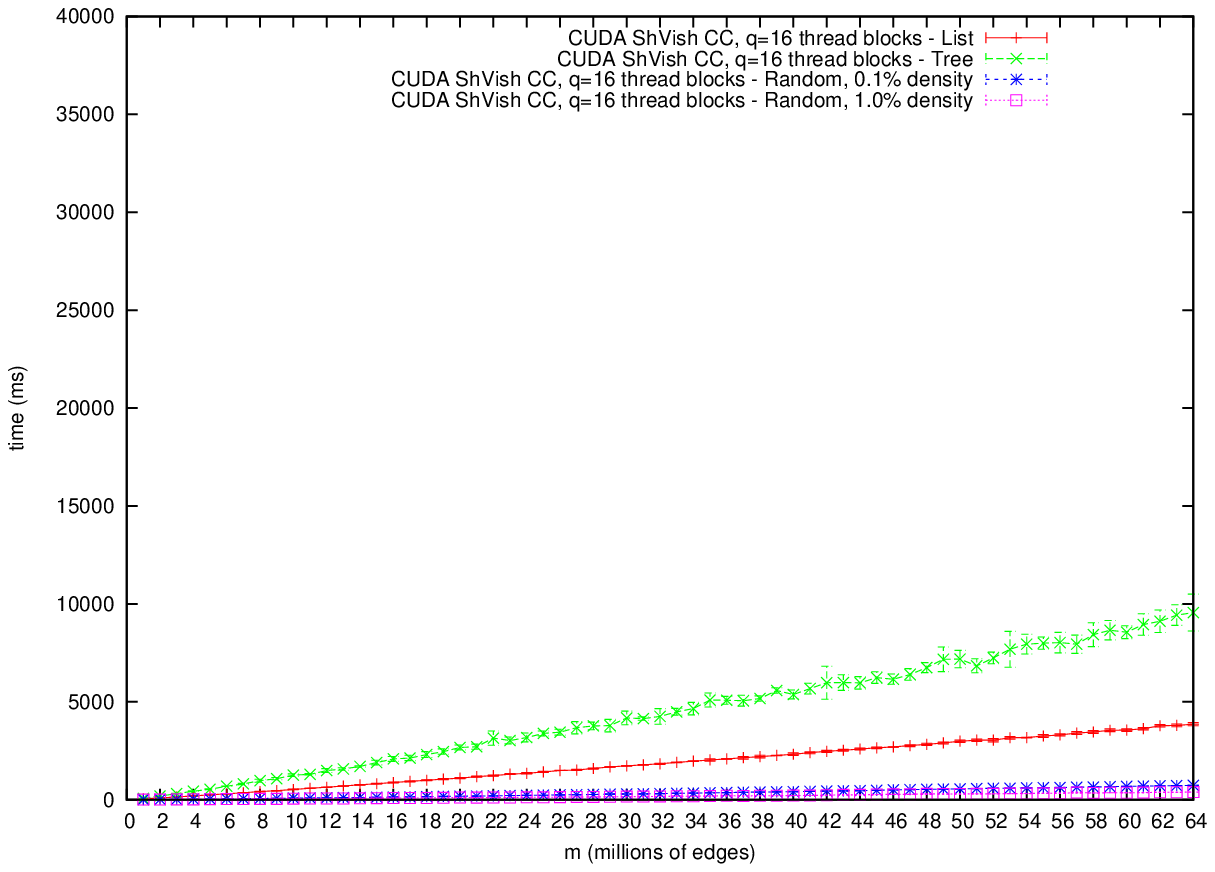}
\par\end{centering}

\caption{Running Times For Sequential and Parallel (GPU) Connected Component
Computation For Different Types Of Graphs\label{fig:Seq-Par-CC}}

\end{figure}

Figure \ref{fig:Seq-Par-CC} shows a comparison of the run times (in
milliseconds), as a function of list size, for sequential and parallel
(GPU) connected component computation for different types of graphs.
Sequential connected component computation was executed on a standard
CPU (Intel Core 2 Quad with 8 GB memory running Fedora Core Linux).
Our GPU implementation of Shiloach and Vishkin's algorithm was executed
on an nVIDIA GeForce 260 with 27 SMs (216 processor cores). Each data
point in Figure \ref{fig:Seq-Par-CC} represents the average of 20
experiments. The experiments use different types of graphs: (1) List
graphs consisting of a collection of random linked lists. (2) Tree
graphs consisting of a collection of random trees of degree $k$.
(3) Random graphs consisting of randomly created connected components
with edge density $d=0.1\%$ or $d=1\%$.

\begin{figure}[H]
\centering{}\includegraphics{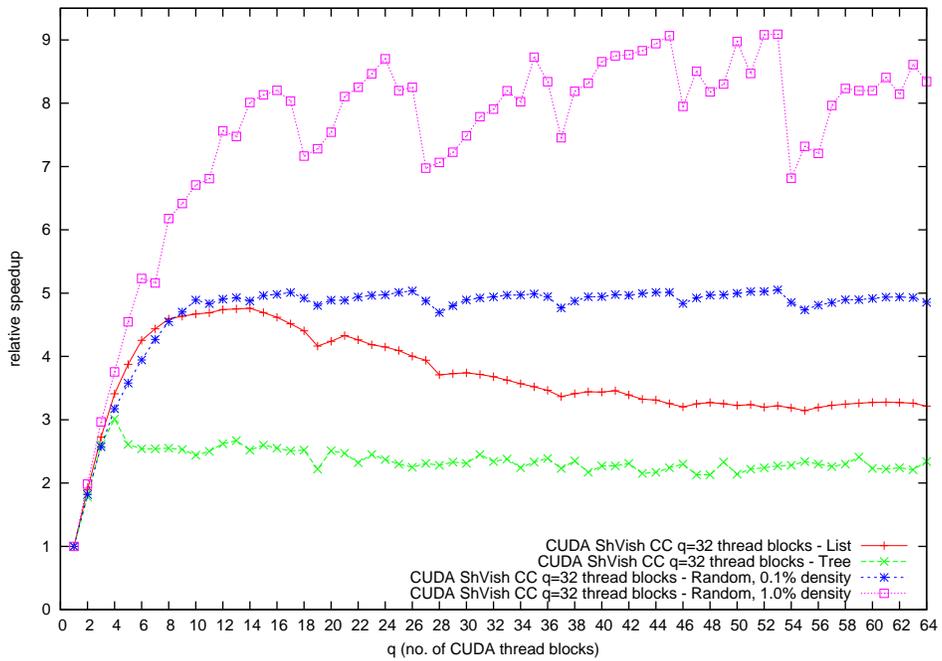}\caption{\label{fig:Comparing-CUDA-CC-Speedup}Comparing CUDA Connected Components
Implementation Relative Speed-up}

\end{figure}

\begin{figure}[H]
\centering{}\includegraphics[width=0.45\columnwidth]{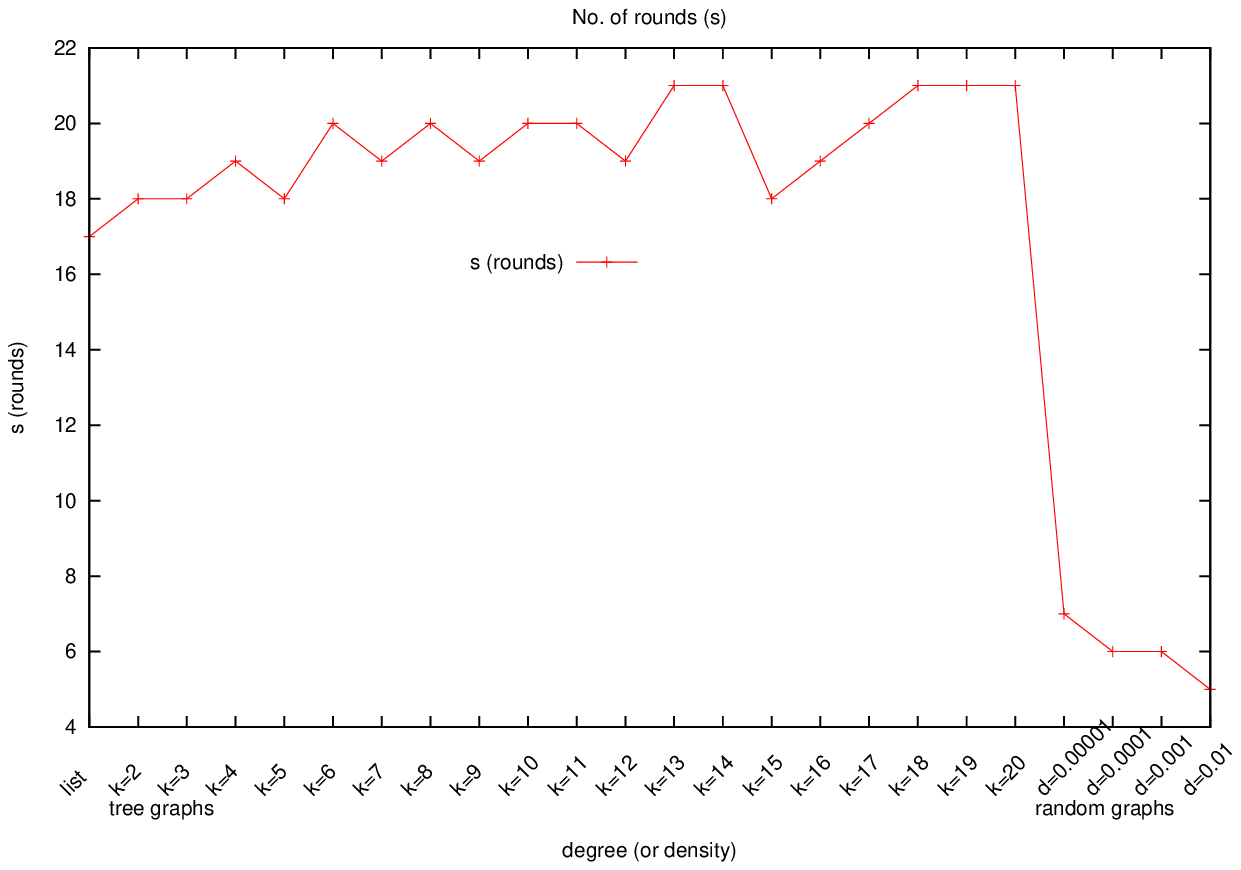}\includegraphics[width=0.5\columnwidth]{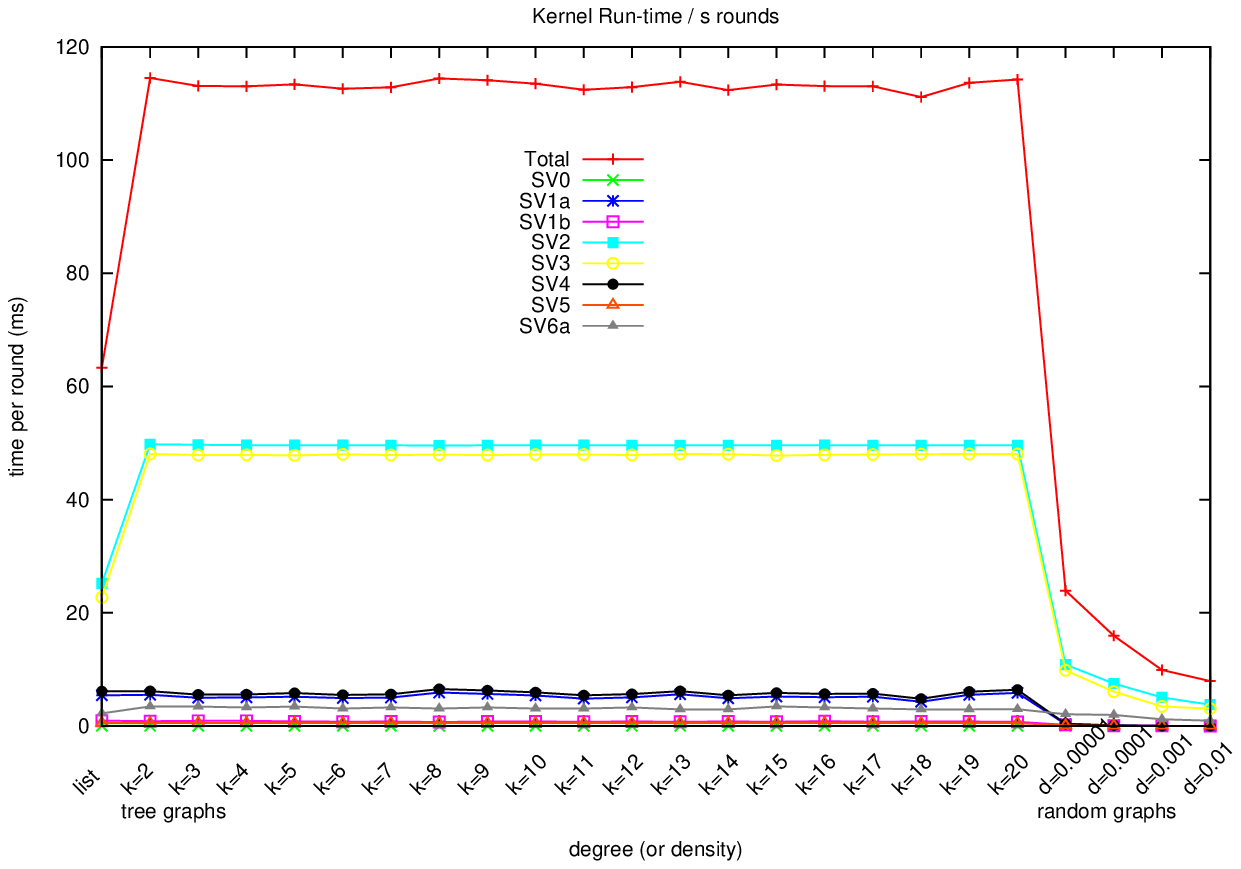}\caption{\label{fig:Comparing-Vary-Deg-On-CCPerf}Detailed Performance Comparison
For Different Types Of Graphs}

\end{figure}

In general, our GPU implementation seems to be successful. Despite
the highly irregular data access patterns which complicate memory
access coalescing, our GPU implementation of Shiloach and Vishkin's
algorithm appears to be considerably faster than the sequential method.
Note that the sequential method requires only linear work while Shiloach
and Vishkin's algorithm requires $O((n+m)\log n)$ work. Another important
observation is that performance is different for different types of
graphs. Here, we will not discuss why this happens for the sequential
methods (e.g. caching effects) and concentrate on our GPU implementation.
Figure \ref{fig:Comparing-CUDA-CC-Speedup} shows in more detail the
performance of our GPU implementation in terms of relative speedup
as a function of the number of thread blocks. Random graphs are processed
more quickly than lists, and lists are processed faster than trees.
All speedup curves show no further improvement for more than 25 thread
blocks which reflects the number of physical SMs available for our
nVIDIA GeForce 260 GPU. Kernels SV0, SV1a, SV1b, SV4 and SV5 of Algorithm
\ref{alg:Shiloach-Vishkin-ConComp-anyP} process vertices and perform
$O\left(n\right)$ work while Kernels SV2 and SV3 process edges and
perform $O\left(m\right)$ work. As shown in \cite{conn-comp-PRAM-OlogN},
the algorithm will iterate at most $\left\lfloor \log_{\frac{3}{2}}n\right\rfloor +2$
rounds but the actual number of rounds will differ according to the
actual graph. The number of actual rounds is in general smaller for
random graphs than for for trees. That explains the better performance
of denser graphs as compared to sparser graphs as compared to trees.
But why do lists show a better speedup than trees? Observe that, inside
Kernels SV2 and SV3 of Algorithm \ref{alg:Shiloach-Vishkin-ConComp-anyP},
for each edge there are 2 and 3 conditions respectively that all need
to succeed to enter the if-block. The failure of any one condition
implies that no work is performed for that edge in those kernels.
The effect on performance is illustrated in Figure \ref{fig:Comparing-Vary-Deg-On-CCPerf}.
The left diagram shows the actual number of rounds for various input
graphs: list graphs (degree $k=1$), tree graphs with degree $k=\left\{ 2,3,\ldots,20\right\} $
and random graphs with density $d=\left\{ 0.001,0.01\right\} $, all
of them with $m=8000000$ edges. The right diagram shows the time
per round spent in each kernel. The time spent for Kernels SV2 and
SV3 dominates the time for the entire Algorithm \ref{alg:Shiloach-Vishkin-ConComp-anyP}.
Kernels SV2 and SV3 do the most work per round for trees, less work
per round for lists and the least work per round for random graphs.
The number of actual rounds is about the same for list and tree graphs
but much less for random graphs which have in general a much smaller
diameter. The total performance is the product of number of rounds
and time per round which explains why the algorithm performs better
for lists than for trees, and best for random graphs.

\newpage

%

\begin{table}[H]
\centering{}\begin{tabular}{|c|c|c|}
\hline 
\textbf{\scriptsize Data Size (for each memory access)} & \textbf{\scriptsize Min. Transaction Size} & \textbf{\scriptsize Segment Size}\tabularnewline
\hline
\hline 
{\scriptsize 1 Byte} & {\scriptsize 16 Bytes} & {\scriptsize 32 Bytes}\tabularnewline
\hline 
{\scriptsize 2 Bytes} & {\scriptsize 32 Bytes} & {\scriptsize 64 Bytes}\tabularnewline
\hline 
{\scriptsize 4 Bytes} & {\scriptsize 64 Bytes} & {\scriptsize 128 Bytes}\tabularnewline
\hline 
{\scriptsize 8 Bytes} & {\scriptsize 64 Bytes} & {\scriptsize 128 Bytes}\tabularnewline
\hline
\end{tabular}\caption{\label{tab:CUDA-1.2-Global-Mem-Segments}CUDA 1.2 Global Memory Segment
Sizes }

\end{table}

\begin{table}[H]
\includegraphics[width=1\columnwidth]{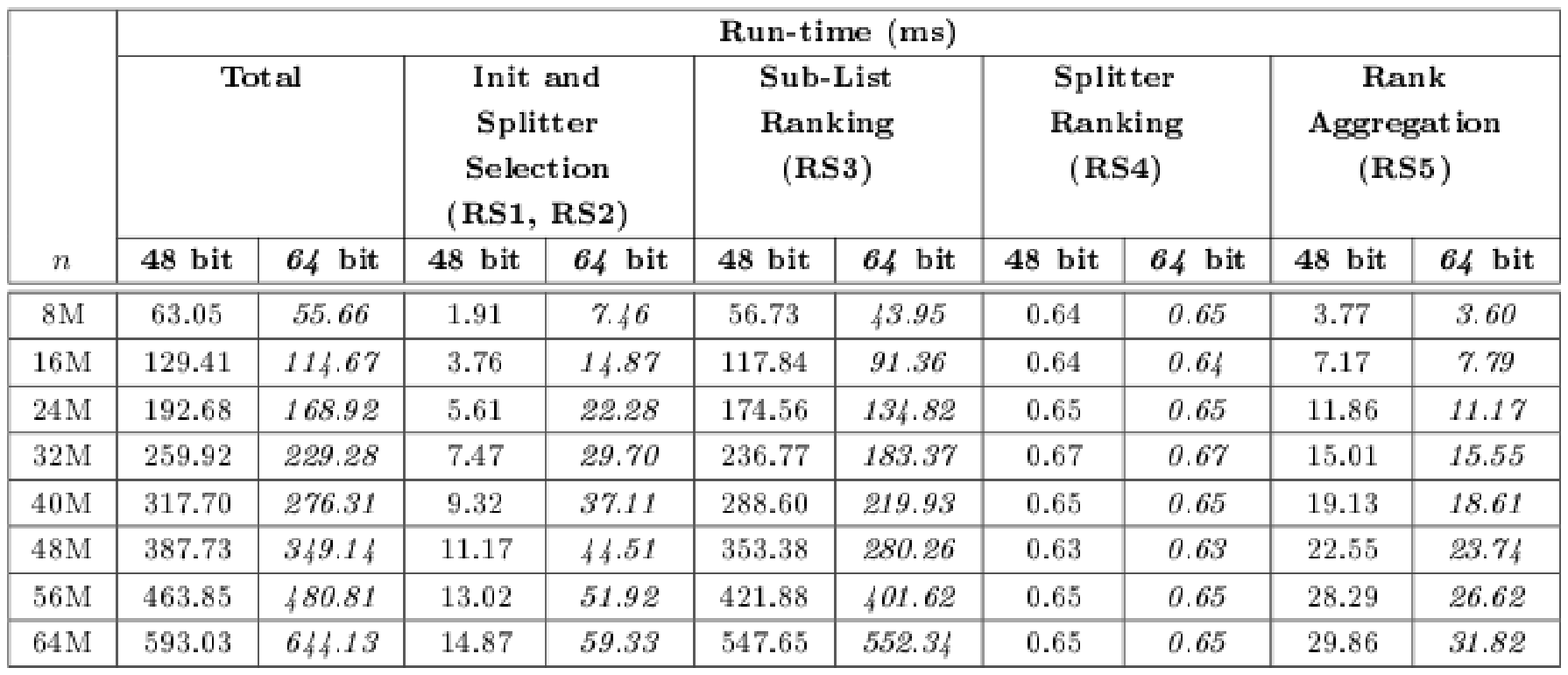}

\caption{\label{tab:Kernel-Timings-LR-ParRandSplit}Kernel Run Times for Parallel
Random Splitter List Ranking, C/CUDA}
\end{table}

\begin{table}[H]
\includegraphics[width=1\columnwidth]{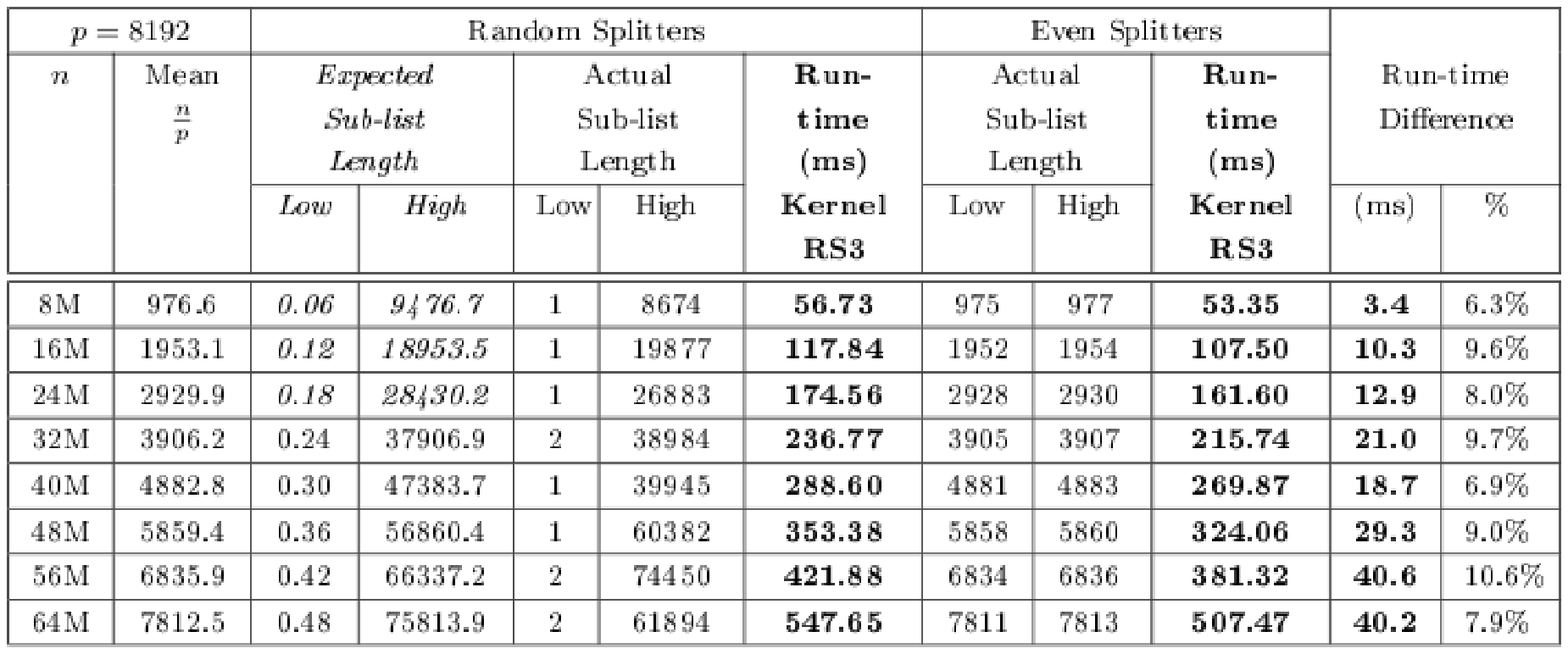}

\caption{\label{tab:Comp-Perf-Random-Even-Splitters}Comparing Kernel Performance
with Random and Even Splitters (48 bit version)}
\end{table}

\begin{table}[H]
\centering{}\begin{tabular}{|c|c|c|c|c|c|c|c|}
\hline 
 & {\footnotesize SV0} & {\footnotesize SV1a} & {\footnotesize SV1b} & {\footnotesize SV2} & {\footnotesize SV3} & {\footnotesize SV4} & {\footnotesize SV5}\tabularnewline
\hline
\hline 
{\footnotesize Work} & {\footnotesize $O\left(n\right)$} & {\footnotesize $O\left(n\right)$} & {\footnotesize $O\left(n\right)$} & {\footnotesize $O\left(m\right)$} & {\footnotesize $O\left(m\right)$} & {\footnotesize $O\left(n\right)$} & {\footnotesize $O\left(n\right)$}\tabularnewline
\hline 
{\footnotesize Reads } & {\footnotesize $0$} & {\footnotesize $2n$} & {\footnotesize $2n$} & {\footnotesize $4m$} & {\footnotesize $5m$} & {\footnotesize $2n$} & {\footnotesize $n$}\tabularnewline
\hline 
{\footnotesize Writes} & {\footnotesize $2n$} & {\footnotesize $n$} & {\footnotesize $n$} & {\footnotesize $2n$} & {\footnotesize $n$} & {\footnotesize $n$} & {\footnotesize $p$}\tabularnewline
\hline
\end{tabular}\caption{\label{tab:Counting-the-RW-In-CCKernels}Counting the Global Reads
and Writes in Connected Component Kernels}

\end{table}

\section{GPU Pseudo Code}

\begin{algorithm}\label{alg:Ptr-Jump-Par-List-Ranking-p<n}
{\bf Wylie's Algorithm On A GPU With $p<n$ Threads (Pseudo Code)}

\includegraphics[width=1\columnwidth]{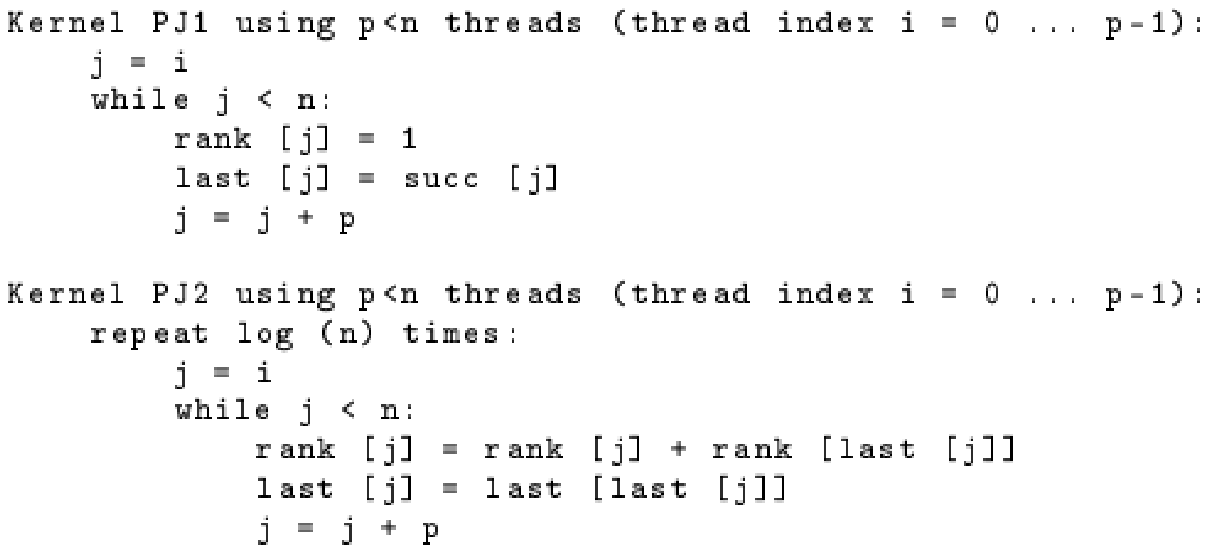}

\end{algorithm}

\begin{algorithm}\label{alg:Rand-split-par-list-rank-p=00003Dr}
{\bf Parallel Random Splitter List Ranking On A GPU With $p=r<n$ Threads (Pseudo Code)}

\includegraphics[width=1\columnwidth]{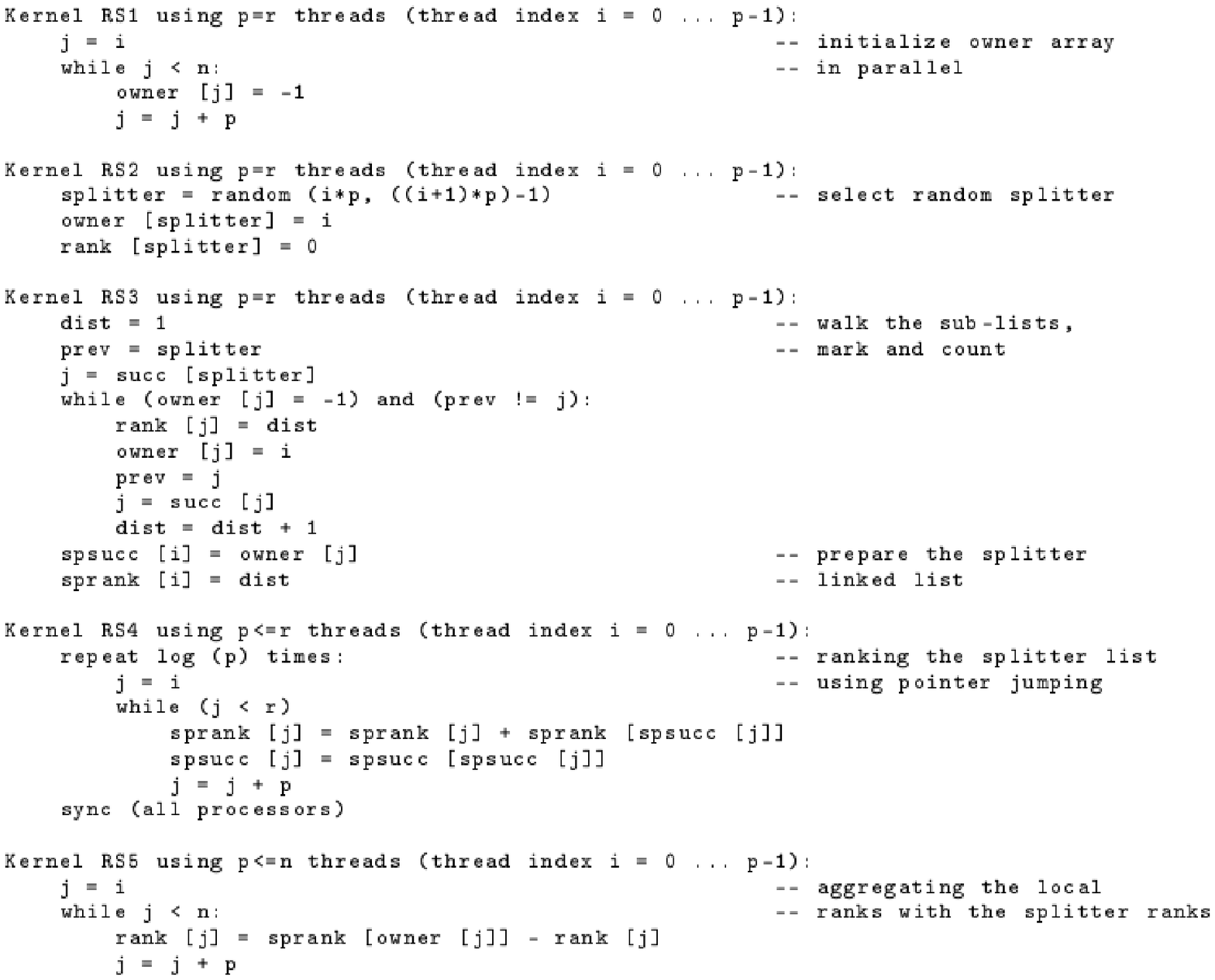}

\end{algorithm}

\newpage
\begin{algorithm}\label{alg:Shiloach-Vishkin-ConComp-anyP}
{\bf GPU Implementation (Pseudo
Code) of Shiloach and Viskhin's PRAM Algorithm with $p\leq n$ Threads
Using Striding. The Step Numbers Indicated Match The Numbering Scheme
Used In \cite{conn-comp-PRAM-OlogN}.}

\includegraphics[width=1\columnwidth]{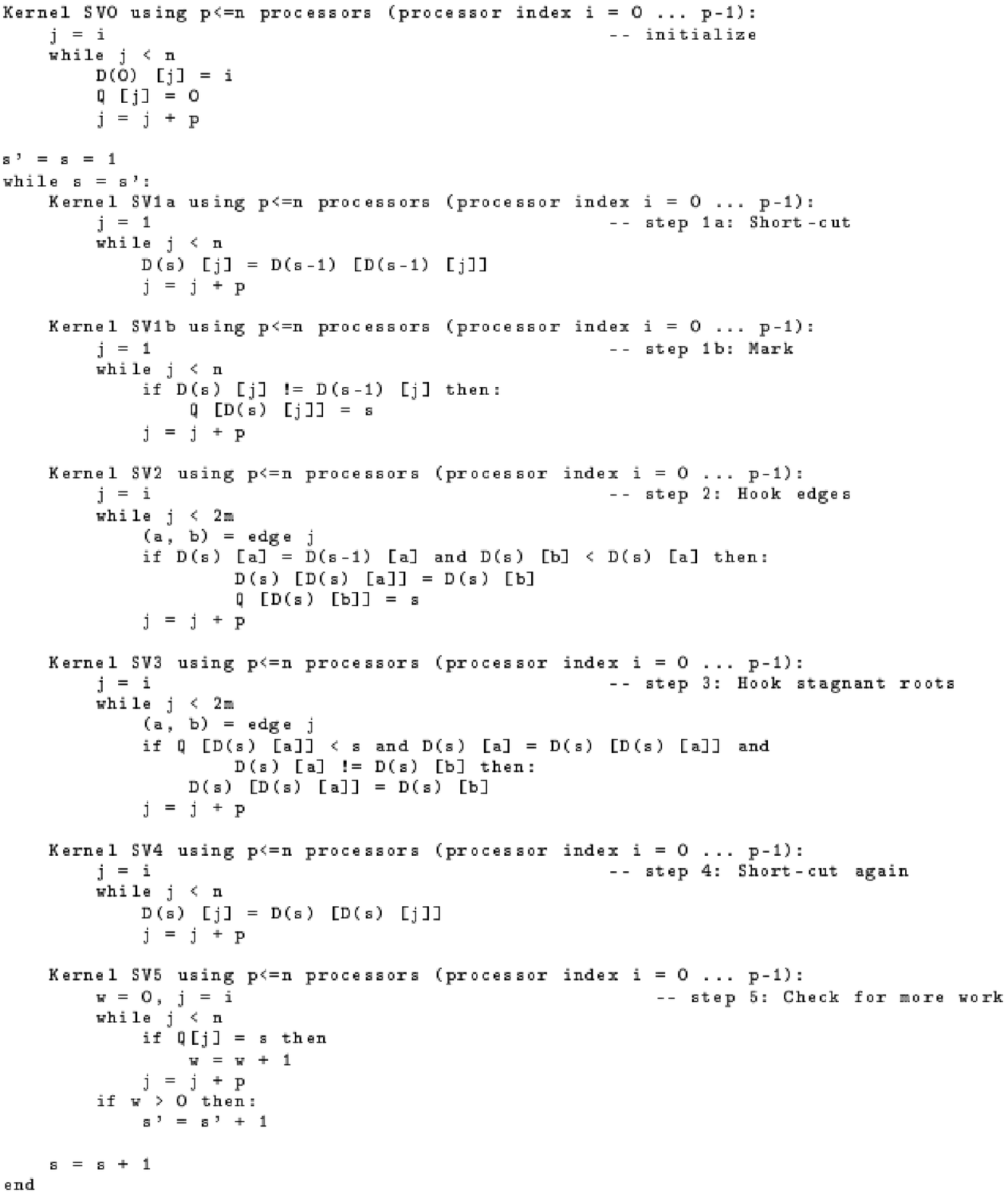}

\end{algorithm}

\end{document}